\documentclass[intlimits,twoside,a4paper]{article}

\usepackage[cp1251]{inputenc}

\usepackage[eqsecnum]{cmpj3}

\usepackage{bm}

\issue{2024}{27}{3}{33803}
\doinumber{10.5488/CMP.27.33803}

\title[Quantifying the role of supernatural entities and the effect of missing data
in Irish sagas]{
Quantifying the role of supernatural entities and the effect of missing data
in Irish sagas}
\author[P. MacCarron]{P. MacCarron\orcid{0000-0002-5163-9264}\refaddr{label1,label2}\thanks{E-mail: \email{padraig.maccarron@ul.ie}.}}
\addresses{
\addr{label1} MACSI, Department of Mathematics and Statistics, University of Limerick, Limerick V94 T9PX, Ireland
\addr{label2} Centre for Fluid and Complex Systems, Coventry University, Coventry, CV1 5FB, England
}

\Keywords{complex networks, social networks, mythology} 

\date{Received January 19, 2024, in final form May 13, 2024}

\begin{document}

\maketitle

\begin{abstract}

For over a decade, complex networks have been applied to mythological texts in order to quantitatively compare them. This has allowed us to identify similarities between texts in different cultures, as well as to quantify the significance of some heroic characters. Analysing a full mythology of a culture requires gathering data from many individual myths which is time consuming and often impractical. In this work, we attempt to bypass this by analysing the network of characters in a dictionary of mythological characters. We show that the top characters identified by different centrality measures are consistent with central figures in the Irish sagas.
Although much of Irish mythology has been lost, we demonstrate that these most central characters are highly robust to a large random removal of edges. 
\printkeywords

\end{abstract}

\section{Introduction}

A key concept in complexity science, is that of \emph{emergence}. This is the idea that seemingly new attributes form in complex systems which are not trivial or expected from analysing their individual constituents. Often these attributes, or properties, can be compared and allow us to classify complex systems, similar to the concept of \emph{universality classes} in statistical physics~\cite{san2023frontiers}. 

Complex networks as an area has recently developed and is often used to analyse complex systems~\cite{Newman2003}. In these systems, the constituents, or \emph{nodes}, are directly linked to one another via \emph{edges}. Structural properties of these networks enable comparisons of different complex systems --- for example one of the first applications compared the neural network of a nematode worm, a power-grid network, and a movie actor network, finding similarities in the structural properties of all three~\cite{watts1998collective}. 

Complex networks have been applied to many systems, from archaeology~\cite{Knappett} to zoology~\cite{DunbarBaboons}. In this work we apply complex networks to mythology. Mythology is one of the furthest fields from science, often rooted in superstition and lying outside the realm of the physical world. However, myths occur in every culture and therefore can be seen as an \emph{emergent property} of a culture. 

Though there are many different definitions of a myth --- a common one being that of a sacred narrative~\cite{Dundes} --- the word itself comes from the Greek word \emph{mythos} meaning \emph{story}, and generally a story concerning the past and the imagination. That is the definition we employ here when using the term \emph{myth}: a traditional narrative from a culture. 

There are many approaches to comparative mythology, such as a structuralist approach to the language~\cite{LeviStrauss}, 
a cognitivist approach~\cite{carney2018psycho},
identifying similar roles and functions of gods in different cultures~\cite{Lyle2006}, and even a phylogenetic approach~\cite{dHuy}, to name a few. Our previous work has looked at comparing the social networks appearing in different myths~\cite{mac2012universal}.

The idea to analyse the social networks of  myths and compare them to one another came from Ralph Kenna. Ralph's interest in universality classes and  complex networks led him to the idea that the myths of different cultures could be compared using their social networks.  The initial question was whether we could classify cultures from the structural signatures of their social networks.
In the end, however, we discovered that different genres of stories gave rise to different  network properties~\cite{mac2014network}. For example, the network of Anglo-Saxon \emph{Beowulf} and the Greek \emph{Odyssey}, both of which feature a central protagonist travelling to different locations, have different network properties to those to the Homer's \emph{Iliad} and the Icelandic \emph{Nj\'als} saga, which feature a larger conflict and are mostly set in one location. The \emph{Iliad} and \emph{Odyssey} are from the same culture and purportedly the same author, yet have different structures.

Ralph was also influenced by Gleiser~(2007)~\cite{Gleiser} which showed why the social network of the comic characters of the Marvel Universe did not look like a real-world social network. Some myths are rooted in history and based on historical events, such as the \emph{Iliad}~\cite{Kraft}. Others are entirely made up, for example, to describe a culture's pantheon or the constellations. However, some myths,  like some of the Irish narratives for example, it is not known where they lie on this spectrum. We hoped to quantify how realistic the social network presented might be, which we attempted in~\cite{mac2012universal}. However, upon identifying that the social network is more reflective of the genre, we 
moved on from this approach in later works.

In this article, a different approach is taken again. We create a large  network of one mythology using a dictionary of characters, and use network centrality measures to identify the important characters. We then randomly and strategically remove edges to determine how robust the identification of these characters is to simulate the effect of loss of texts. This kind of approach has not been applied in this context before.

Many sources of myths and epics have been lost, for example some Greek epics are known in name only~\cite{morford1999classical}. Similarly, there are named Irish narratives in manuscripts which we no longer have~\cite{williams2017ireland}. Estimating how many narratives remaining is also not trivial. Kestemont et al.~\cite{kestemont2022forgotten} use an unseen species model on surviving texts in manuscripts to show that Irish texts could have up to 80\% survivability. However, they do not (and likely cannot) distinguish between texts with narratives or scripture and gospel. The Irish stories that survive 
have a long oral tradition, often being transmitted for centuries before being recorded mostly in the 12th -- 14th centuries~\cite{nagy1986orality,croinin2016early}. 
As the supernatural characters turn up in different sagas, this  allows us to create one large social network.
Due to this, they will be important in this network even if we are missing information on them.
Hence, a seemingly minor character appearing in a few texts, could have been important in the entire mythology but their story is now lost and their network properties will allow us to identify this. 

In this work we focus specifically on early-modern and medieval Irish narratives. 
Traditionally this entire corpus is separated into four \emph{cycles}: the mythological cycle, the Ulster cycle, the Fenian cycle and the historical cycle. The first of these is to do with the settling of Ireland and the supernatural \emph{Tuatha D\'e Danann} (``the tribe of the goddess Danu'').
While some characters of the Tuatha D\'e can be called (or associated with) gods, 
they can be killed. These characters have magical powers, and sometimes are associated with things like the sea (Manann\'an Mac Lir), death (the M\'orrigan), love (Aonghus \'Og), etc. 
Importantly, there is no knowledge or discussion of religious practice to these in the recorded texts making it more challenging to assert whether they are gods or not~\cite{williams2017ireland}.

At the end of the mythological cycle, the Tuatha D\'e retreat underground to the ``Otherworld'' and frequently appear in the other cycles. Chronologically the next cycle is the Ulster cycle. Ulster is a province in the north part of the island. This cycle is centred on the boy-hero C\'uchulainn, the son of the Tuatha D\'e Danann character Lugh (sometimes identified as one of the most important gods~\cite{williams2017ireland}). The following cycle is the Fenian cycle, focussed on the hero Fionn Mac Cumhail and his band of warriors, the Fianna. These myths tend to take place over the whole Ireland though Fionn's base is on the Hill of Allen in Leinster in the east of the island. Finally the historical cycle often deals with voyages from Ireland that tend to be set in the time of historical High Kings of Ireland. 

The classification of myths into four cycles  is rather problematic, however. For example, some of the sagas dealing with the Tuatha D\'e characters span hundreds of years, sometimes taking place across the different cycles, and others are more recently written~\cite{williams2017ireland}. Reference \cite{carey} argues that there is no real separation between the sagas of Mythological and Ulster cycles.
The dictionary used here often does not attempt to categorise a character into one of these cycles, instead, we aim to do this from the network.

One of the disadvantages of our previous methods is the fact that it requires gathering every interaction between characters in texts. For single narratives, this is quite achievable, but for an entire corpus of narratives, this requires a lot of time and manpower. While work has been done on using natural language processing to extract characters and character networks~\cite{labatut2019extraction}, it works best for extracting conversation links and 
more recent varying attempts do not achieve huge success~\cite{maddy2023}. 
Here, we attempt to bypass this by using a dictionary of mythological characters and scraping the interactions from other names mentioned in their entries. We  use centrality measures and inter-community edges to attempt to quantify the most important  characters. We then randomly and strategically remove edges, to represent missing connections due to lost narratives, and categorise how stable the central characters are.

This paper is structured as follows: in the next section the data and methods used are described. The following section give the network properties, identifies the central characters, analyses the communities and performs link removal simulations for the robustness. Then, the conclusions and plans for next steps are presented.

\section{Data and methods}

We use Peter Berresford's \emph{Dictionary of Irish Mythology}~\cite{ellis1991dictionary} to obtain data on Irish mythology. This contains entries on 776 characters from Irish myths. There are a further almost 200 characters mentioned in descriptions that do not have their own entries. As many characters have the same names, and this contains information on place names (and sometimes academics who have worked on that character), much of the data had to be evaluated by hand after scraping to ensure accuracy.

A directed edge is added each time a character is mentioned in the entry for another character.  No further meaning is ascribed to an edge. Therefore, each character will have an \emph{in-degree} (representing the number of times they appear in other character's entries), and an \emph{out-degree} (which represents how many other characters are mentioned in their description). We will fit to these distributions using the methods described in reference~\cite{mannion2023robust}. 

Beyond the degree, we will determine the most central characters in the network using the betweenness centrality which identifies characters on many shortest paths~\cite{Freeman}. Therefore, if information propagates through the network, it is likely to pass through characters with a high betweenness. We also use the closeness centrality which is the reciprocal of the sum of shortest distance (i.e., farness) to all other nodes~\cite{bavelas1950communication}. 

To detect communities, we use the Girvan-Newman algorithm~\cite{GirvanNewman}. 
This works by recursively removing the edge with the highest betweenness centrality until it finds a partition. 
We will use the modularity to verify that the partition is better than a partition selected at random~\cite{NewmanGirvan}. 
While modularity in community-detection algorithms has its downfalls~\cite{fortunato2007resolution,peixoto2021descriptive}, here we are using it in a descriptive capacity rather than an inferential one. Also, as we have the distinct advantage here of 
knowing some of the members of the communities, we can discern if the communities detected describe the system appropriately or not.

\begin{figure}
\centerline{\includegraphics[width=1.0\textwidth]{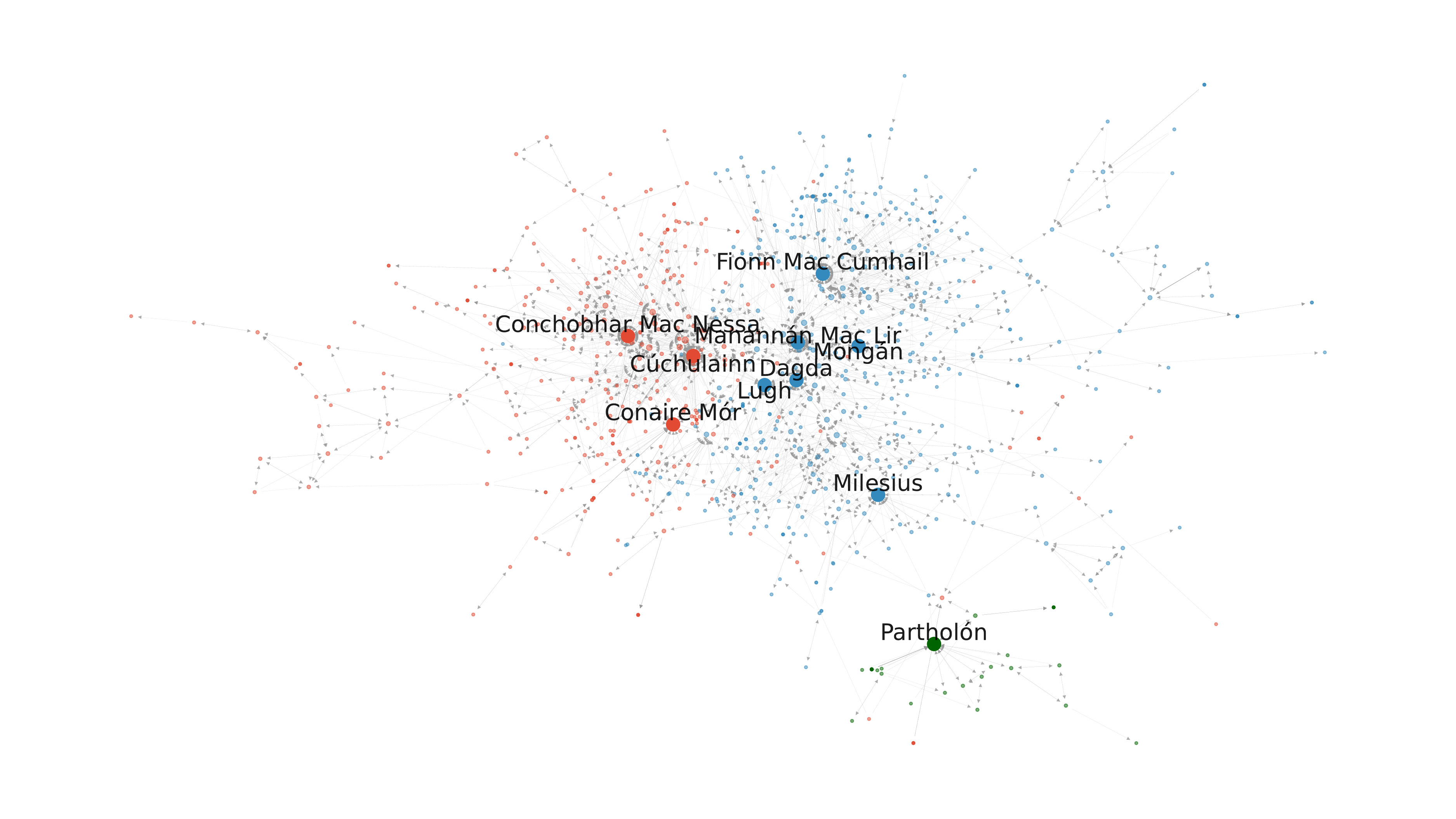}}
\vspace{-2em}
\caption{(Colour online) The directed network with 3 communities fitted using the Girvan-Newman community-detection algorithm. The ten characters with the highest betweenness are named. Of these, four are supernatural characters, two are kings, two are leaders of the settlers of Ireland and the others are the two major heroes.} \label{fig:net}
\end{figure}

With the communities identified, we will identify inter-community edges and identify frequent characters connecting communities. In the Irish data, we expect these characters to be of the Tuatha D\'e Danann.

Finally, we perform a simulation to test the effects of missing edges. As mentioned earlier, many of the sources have been lost. Hence, we wish to see how robust the central characters are if we remove edges at random and strategically. 
We use the Jaccard index for the top 5, 10 and 25 characters by both betweenness and closeness. The Jaccard index does not care about the order, just that the set is the same. We perform a random removal of 500 edges 1,000 times and compute the 2.5 and 97.5 percentiles from the 1,000 simulations to get a 95\% confidence interval around the mean. We then take a strategic removal approach related to snowball sampling. This takes the idea that missing edges are unlikely to be random, but related to the characters interacting in specific stories. Here, we randomly pick a character and with probability $p$ remove edges between each of their neighbours. We then go to each neighbour and with probability $p-n\epsilon$ (where $n$ is the distance from the original node and $\epsilon$ is a small value) remove an edge for each of their neighbours. We recursively repeat this until $p-n\epsilon \leqslant 0$. Each time an edge is removed, we compute the Jaccard index again on the betweenness centrality and repeat the process until 500 edges are removed. This is also repeated 1,000 times. We choose $p=0.8$ and $\epsilon = 0.2$.

\section{Results}
\subsection{Core network properties}

The weakly connected giant component of the network is shown in figure~\ref{fig:net}. This comprises 768 of the 959 named characters. Of these, 525 are male, 175 are female, and the rest are unknown. There are 2,164 edges giving an average degree of 2.26. The reciprocity of the directed network is 0.46 indicating that less than half the edges go both ways.

\begin{figure}[!b]
\includegraphics[width=0.49\columnwidth]{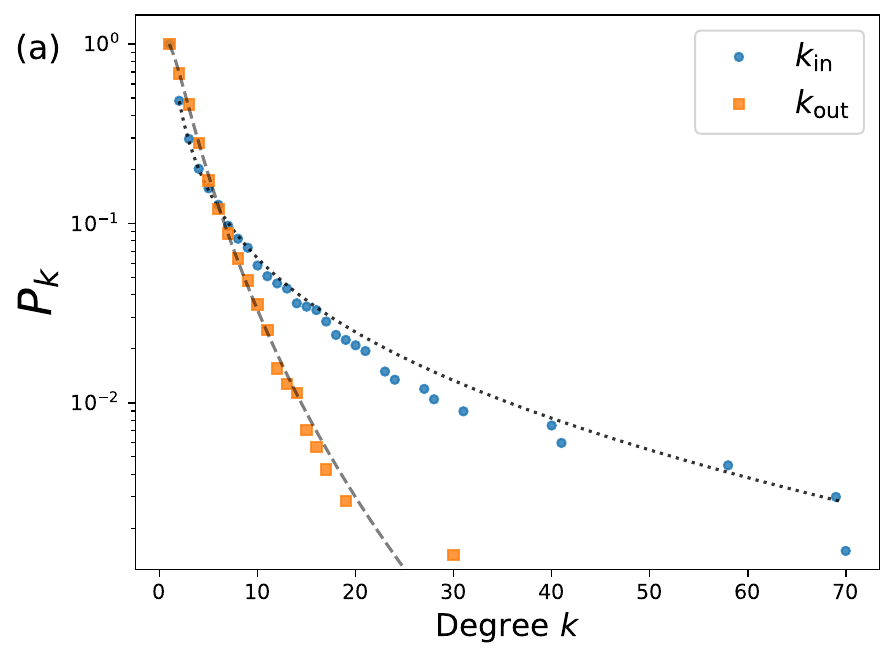}
\includegraphics[width=0.49\columnwidth]{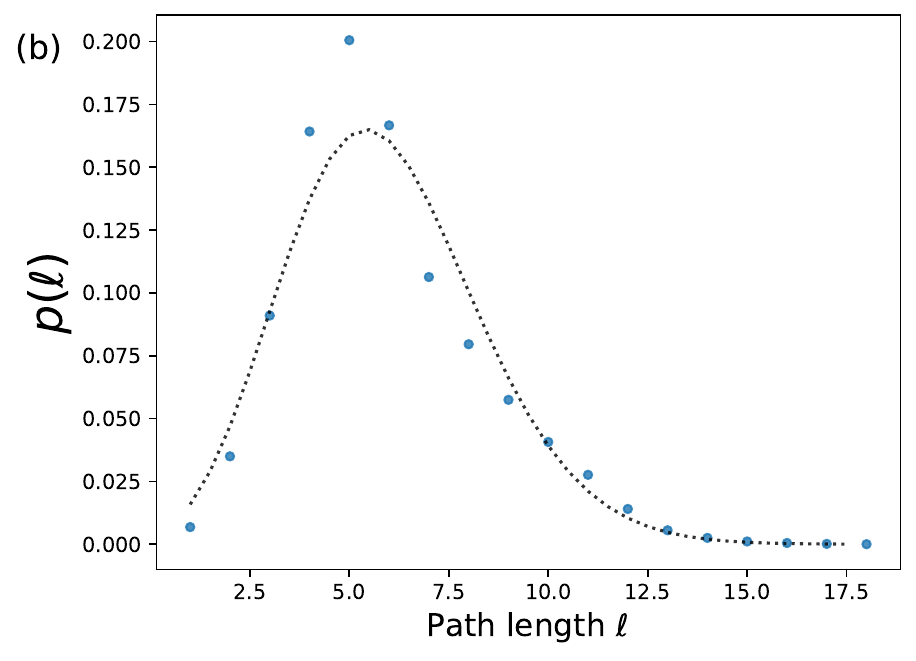}
\caption{(Colour online) Panel (a) shows the complementary cumulative degree distribution $P_k$ for the in-degree (circles) with a fitted truncated power law, and out-degree (squares) with a fitted log-normal distribution. Panel (b) shows the probability mass function for the shortest path length $p_\ell$ with a fitted Poisson distribution.} \label{fig:dist}
\end{figure}

The in-degree and out-degree distributions are shown in figure~\ref{fig:dist}~(a). Fits are made to the distributions using the methods outlined in~\cite{mannion2023robust}. Of the candidate distributions, the in-degree distribution receives most support from a discreetised power law with an exponential cut-off of form $p_k\sim k^{-\gamma} \re^{-k/\kappa},$ with parameter estimates of $\gamma=1.97$ and $\kappa=115.04$. There is a moderate support for a power law with exponent 2.14. The out-degree distribution receives best support for a log-normal distribution of form $p_k \sim (1/k) \exp[-{(\ln k - \mu)^2}/{2\sigma^2}]$ 
with parameter estimates of  $\mu = 0.80$ and $\sigma=0.79$. Moderate support is received for an exponential distribution of form $p_k \sim \exp(-k/\kappa)$ with parameter estimate $\kappa=2.53$. Since the out-degree decays much faster and is confined to an individual entry, we attribute more significance to the in-degree as this requires characters frequently mentioned in many descriptions. 

We show the top 20 characters ordered by in-degree and out-degree in table~\ref{tbl:cent}.
Both C\'uchulainn and Fionn Mac Cumhail, the two heroes the Ulster cycle and Fenian cycles are centred on respectively, have the highest in-degrees (70 and 69) and out-degrees (30 and 19). As evident from figure~\ref{fig:dist}~(a), these in-degrees are far to the right of other characters. The supernatural Lugh L\'amhfhada has the third highest out-degree (17) but 12th highest in-degree (21). While the in-degrees and out-degrees are correlated $r=0.69$ ($p<0.001$), when ranked they are not strongly correlated with a Spearman-rank correlation of $\rho=0.44$ ($p<0.001$) reflecting the different ordering of characters. The degree assortativities of the graph are uncorrelated with an out-out and in-in assortativities of $-0.01$ and $-0.04$, respectively.

\begin{table}
\caption{The top 20 characters arranged by in-degree, out-degree, betweenness centrality and closeness centrality. The values of the in-degree and out-degree are given in parenthesis. Character names in bold are supernatural characters from the Tuatha D\'e Danann.} 
\label{tbl:cent}
\vspace{2ex}
\begin{center}
\begin{tabular}{|c|c|c|c|c|}
\hline
   Rank  & In-degree&   Out-degree& Betweenness & Closeness\\
\hline \hline
1 & C\'uchulainn (70) & C\'uchulainn (30) & C\'uchulainn & \textbf{Manann\'an} \\
2 & Fionn (69)& Fionn (19)& \textbf{Manann\'an} & C\'uchulainn \\
3 & Conchobhar (58)& \textbf{Lugh} (17)& \textbf{Dagda} & \textbf{Dagda} \\
4 & Medb (41) & Naoise (16) & \textbf{Lugh} & Medb \\
5 & \textbf{Manann\'an} (40) & Oscar (15) & Fionn  & Ailill \\
6 & Conall Cearnach (31)& Conchobhar (14)& Milesius & \textbf{Lugh} \\
7 & Parthol\'on (28)& Fergus (14)& Conchobhar & \textbf{Bodb Dearg} \\
8 & Ailill (27)& Milesius (14) & Conaire M\'or & Cormac Mac Art \\
9 & \textbf{Aonghus \'Og}(24)& Deirdre (13)& Parthol\'on & Fionn  \\
10 & \textbf{Dagda} (23)& Diarmuid (12) & \textbf{Mong\'an} & \textbf{Aonghus \'Og} \\
11 & \textbf{Bodb Dearg} (21)& Ois\'in (12) & \textbf{Ogma} & \textbf{Fand} \\
12 & Cormac Mac Art (21) &\textbf{ Aonghus \'Og} (11)& \textbf{Midir} & \textbf{Nuada} \\
13 & \textbf{Lugh} (21) & Conaire M\'or (11)& Eber  & \textbf{Midir} \\
14 & \textbf{Nuada} (2)& \'Etain (11)  & Ron\'an & Conall Cearnach \\
15 & Naoise (19)& \textbf{Ethlinn} (11)& \textbf{Bodb Dearg} & Conchobhar \\
16 & Fergus (18) &\textbf{ Manann\'an}(11)& Dubhthach & Balor \\
17 & Conaire M\'or (17)& Niall No\'ighiallach (11) & Fiacha  & Conaire M\'or \\
18 & Milesius (17)& \textbf{Ogma} (11)& Diarmuid  & Dechtir\'e \\
19 & \textbf{Ogma} (17) & Art (1)& Aedh Dubh & \textbf{M\'orr\'ig\'an} \\
20 & Tuireann (17) & Balor (10) & \textbf{Aonghus \'Og }&\textbf{ B\'ecuma} \\
\hline
\end{tabular}
\renewcommand{\arraystretch}{1}
\end{center}
\end{table}

 The mean number of steps between two characters is 5.92 ($\sigma=2.41$) with a maximum shortest distance of 18. The full path-length distribution is shown in figure~\ref{fig:dist}~(b) with a fitted 
Poisson distribution ($\lambda=5.92$) capturing the right skew. Note that for the shortest distance, we convert the graph to undirected, but all other measures use directed counterparts (including the distances in the betweenness and closeness centralities below).

Instead of focussing on the number of connections, however, we are more interested in where they lie on the network. We use the betweenness and closeness centralities as a more representative measure of influence. The higher is the betweenness, the shorter are the paths between other pairs a node lies on. The higher is the closeness, the less distance it has to all other nodes.

The top 20 nodes ordered by betweenness and closeness centrality are displayed in table~\ref{tbl:cent}. The order is different again to the in-degree and out-degree. The closeness centralities contain more of the supernatural characters (in bold), ten in total. Closeness is also the only to feature one of these characters, Manann\'an Mac Lir (sometimes referred to as the god of the sea), at the top.  
We also compute the eigenvector centrality (not shown in table).

The ten nodes with the highest betweenness centrality are named in the network diagram in figure~\ref{fig:net}. Two of these are the main heroes of the Ulster and Fenian cycles, C\'uchulainn and Fionn. Conchobor Mac Nessa is the king of Ulster in C\'uchulainn's time and Conaire M\'or is the high king of Ireland at that time. The first cycle tells us of seven settlings of Ireland. One of the first is that of Parthol\'on, athough they all die of a plague. Hence, they are not strongly connected to the main network and as a result Parthal\'on's betweenness is inflated. After the Tuatha D\'e Danann settle,  the Milesians, led by Milesius, arrive next and drive the Tuatha underground. The remaining four characters named are all influential members of the Tuatha.

In terms of order, C\'uchulainn has the highest betwenness, followed by Manann\'an Mac Lir, the Dagda and Lugh before Fionn. The Dagda is the original head of the Tuatha D\'e Danann. Lugh and Mannan\'an are the two who tend to appear more frequently in different narratives usually to aid the protagonists. While it is not surprising to see these rank highly, it is surprising to see Mong\'an, the son of Manann\'an Mac Lir in the top ten. He is not a character of major significance except in one narrative. However, he links characters in a few different narratives increasing his importance in the network.

Using the eigenvector centrality yields surprisingly different results, of the top ten characters here (Conchobhar Mac Nessa, C\'uchulainn, Naoise, Conall Cearnach, Cathbad, Deirdre, Medb, Fergus Mac Roech, Ailill Mac M\'ata, and Manann\'an Mac Lir), 9 of those are all in the Ulster cycle (in the red cluster in figure~\ref{fig:net}). This seems to identify the core characters of one cluster, with only one supernatural character in the top 20 and Fionn in fifteenth. This is similar to the recent work on a  Russian novel which shows that betweenness centrality corresponds well to characters that are structurally important, and eigenvector centrality to the most important sub-network~\cite{tarasevich2023network}.

\subsection{Communities}

The Girvan-Newman algorithm is used to determine communities in the giant component~\cite{GirvanNewman}. The first partition separates the people of Parthol\'on, 37 nodes ($Q=0.05$). After this there is a split to 426 and 309 nodes. The modularity of these three clusters jumps to $Q=0.48$. If we continue partitioning, the algorithm continues to remove clusters of 20 or 30 nodes but does not split the larger communities substantially. The modularity slowly grows and does not jump again like from two communities to three. Hence, we settle on three communities which are coloured in figure~\ref{fig:net}.

Inspecting the membership of the two large communities, we see the larger one (blue in figure~\ref{fig:net}) contains the Fenian cycle and characters from historical cycle. The smaller one (red) contains characters from the Ulster cycle. The characters of the Tuatha D\'e appear as frequently in the other cycles as the mythological one so they are spread throughout both clusters. We find two clear misclassifications, the brown bull Donn C\'uailnge, the chief prize of the epic \emph{T\'ain B\'o C\'uailnge}, and the Ulster cycle character Fraoch are both in the blue rather than red clusters. This is because Bodb Dearg is mentioned in Donn C\'uailnge's entry and Fraoch is the son and nephew of the Tuatha characters B\'e Find and Boann, respectively. Bodb Dearg and Boann both have strong connections to characters in the Fenian cycle, the former being the head of the Tuatha D\'e at that time. 

Of the 54 edges between the two largest communities, half of them contain a character from the Tuatha D\'e Danann. Some of these such as Aonghus \'Og and the M\'orrig\'an are well known, characters such as Ogma and  Midir, however, are less known. Many of these characters appear in other narratives but are almost never central characters in any narrative.

\begin{figure}[!t]
\includegraphics[width=0.495\columnwidth]{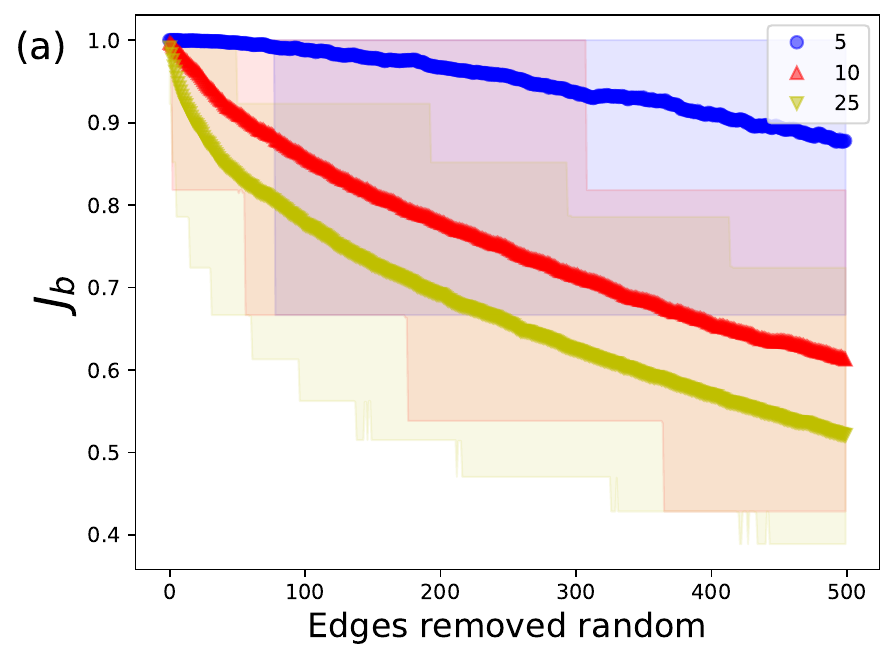}
\includegraphics[width=0.495\columnwidth]{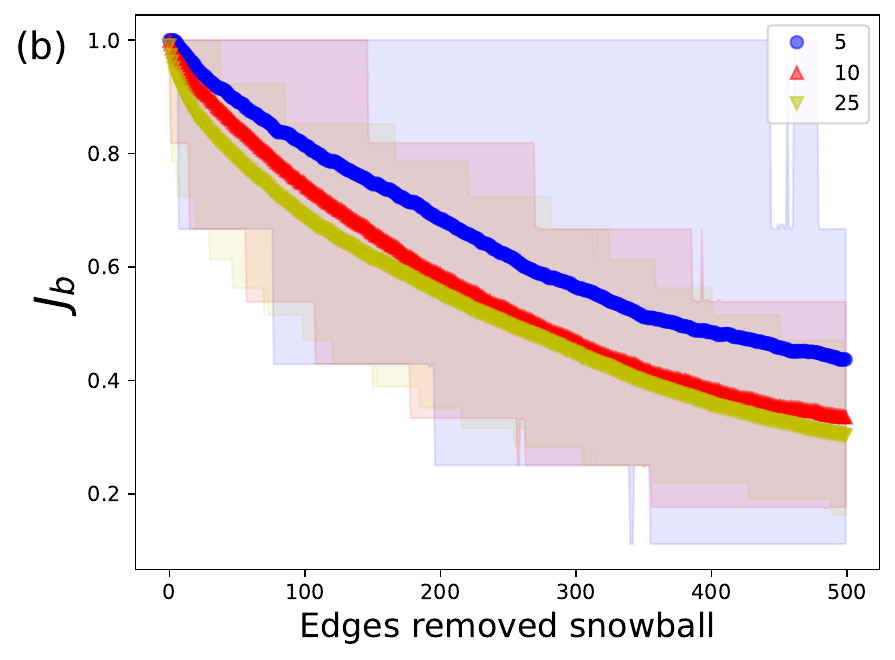}
\caption{(Colour online) Panel (a) displays the Jaccard index for the top 5, 10 and 25 nodes ranked by betweenness $J_b$, here we see even removing 500 (almost 25\%) of the edges, the Jaccard stays above 0.5 showing that more than half the characters in the top 25 do not change. Panel (b) shows the Jaccard index when removing characters strategically is not as robust. } \label{fig:sim}
\end{figure}

\subsection{Simulations}

As mentioned earlier, much of Irish mythology has been lost. In order to measure the effect that the missing edges have, we randomly and strategically remove the edges to determine if the central characters are consistent. 
We measure this by checking the 
Jaccard index for the top 5, 10 and 25 most central characters. 
We initially randomly remove an edge and compute these 1,000 times up to the removal of 500 edges. We plot the mean surrounded by the 2.5 and 97.5 percentile as a 95\% confidence interval. We then strategically remove the same number of edges, by randomly selecting a node and removing with probability $p=0.8$ their neighbour's edges. We then go to their neighbours and remove edges of those with $p-n\epsilon$ where $n$ is the distance for the first node and we chose $\epsilon=0.2$. This reflects the idea that these characters are all in the same story.

In figure~\ref{fig:sim}~(a) 
we show the Jaccard index (the size of the intersection of two sets over the size of the union) between the top ten characters ranked by betwenneness compared to the top 5, 10 and 25 nodes after each edge removal. Here, we see that even after removal of almost 25\% of the edges, the Jaccard index remains above 0.5 indicating that half of those characters remain in the top 25 at worst. 
Note, the lack of change on the quantiles shows that certain regions are highly stable. 
Similar results are found for the closeness centrality (not shown).

Focussing on the top five characters based on betweenness, we randomly remove 500 edges 1,000 times. C\'uchulainn, Manann\'an, Lugh and Fionn remain in the top 5 100\% of the time, Dagda remains 96\% of the time. Going to the next five, however, Conchobhar remained 98\% of the time, but the other four fluctuate more which is why the Jaccard index approaches 0.6 in figure~\ref{fig:sim}(a). Characters such as the M\'orrig\'an, Ogma, Aonghus \'Og, and Midir show up in the top ten frequently at these points.

Figure~\ref{fig:sim}~(b) shows the Jaccard index for the
 strategic removal of nodes. Here, we do not have the same level of robustness. After removing almost 25\% of the edges, the Jaccard index is less than 0.5 for the top 10 and top 25 nodes meaning that less than half of them are the same.

From this, we conclude that  with a random loss of up to a quarter of the edges, the top five characters are highly stable when ranked by betweenness and closeness. Going beyond this, there is more variation, but all from characters who are identified in table~\ref{tbl:cent}.
In this case, with the random loss of edges,
the central characters are likely to be consistent when removing randomly at least. However, when taking a snowball sampling approach, the central characters are less robust. However, the parameter space here is quite large and further investigation needs to be done for this. As many of the main characters frequently co-occur in a given cycle, perhaps a value of $p=0.8$ is too high. Future work from the individual texts is required to ascertain good parameter values.

\section{Conclusions}

When studying an amalgamated network of 18  sagas of the Icelanders, we observed that the highest betweenness characters were generally not the protagonists of single narratives~\cite{mac2013network}. Some chieftains like Snorri Go{\dh}i or Grimr, 
feature highly, as does a relatively minor character named Olaf the Peacock who was a merchant who then became a chieftain.  
That work featured over 1.500 characters. It was this realisation that characters do not need to be central to a text but can have large influence (possibly related to their status) in the combined corpus that led us to the aim of quantifying the role of gods in a mythology.

This led us to the subsequent question: who are the most important characters in Irish mythology? 
The author of the dictionary used seems to argue that it is Lugh and calls him a sun god (though there is little evidence to support this~\cite{williams2017ireland}). Others argue that it is Dagda, the head of the Tuatha D\'e Danann. There is merit for it to be Manann\'an Mac Lir, often called god of the sea, who is the earliest pagan deity named in the narratives~\cite{williams2017ireland}. From our directed network, 
we quantitatively demonstrate that these three are most important Tuatha D\'e characters for the network of dictionary characters. At the very least, this tells us that these data are reliable. At best, this is support for their influence in the Irish sagas.
Unlike the two main heroes C\'uchulainn and Fionn, these are generally not the protagonists of many narratives. However, if information were to pass through the network, it would be more likely to pass through one of these characters. 

These characters also feature on many inter-community edges between the two cycles with the most narratives. They continue to feature highly in the network properties with the random removal of up to 25\% of the edges.
This implies that even if we only have a fraction of narratives from that time, it is likely that these three characters would still be the core supernatural characters in Irish myths.  Though, when we take a non-random approach, there is less stability in the central characters, the choice of parameters here may reflect that and this needs to be investigated further.

Earlier the issues of traditional classification into four cycles was mentioned. From the network, we can easily detect two large communities, one focussed on C\'uchulainn and the other on Fionn. This supports the idea that the Mythological Cycle should not be separate as argued qualitatively in reference~\cite{carey}.

It is worth noting that the network here does not represent the same type of network we have previously analysed. In those we have edges each time two characters interact in a story. In the dictionary we do not have as precise edges, sometimes another character is mentioned if an event they are associated with is similar to that of the discussed character. Hence, we use directed edges here because that is the only further information we have. This has implications for the betweenness calculations which affects both our character-ranking and community-detection methods. However, our simulations show that the order of characters by this is robust for the major characters.

Going forward, we aim at continuing gathering data for individual myths and hope to complete all the Irish narratives. This will allow us to validate the results here on a much larger network and get more accurate results for the loss of edges and test the loss of random narratives. 

The methods presented here have just been focussed on Irish mythology. However, this could easily be applied to any other mythology where a similar dictionary is available. As mentioned earlier, some Greek epics are known to have existed that we no longer have, the edge-removal technique here could also be used in those data  to quantify the change in central characters there in order to simulate and measure the effect of loss of data. Dictionary data also exist more commonly for other cultures, and datasets like Wikipedia could be used to gather links similar to the methods employed by reference~\cite{schwartz2021complex}. This would speed up the data collection, although for the Irish data it is a bit lacking at present.

\section*{Acknowledgements}

I would like to thank and dedicate this work to Ralph Kenna, who, without his creativity and ability to see connections between fields, this work could never have been realised. The first project I began with Ralph in 2010 involved gathering these dictionary data. In the end, we moved to individual narratives and did not return to these data but always had the intention to. He would have appreciated the final results and it is a great shame he was not involved in finishing this. He is dearly missed for his insight, guidance, and friendship.

\bibliographystyle{cmpj}
\bibliography{bibliography}
\newpage
\ukrainianpart

\title{Кількісна оцінка ролі надприродних сутностей і впливу відсутніх даних в ірландських сагах}
\author{П. Маккаррон\refaddr{label1,label2}}
\addresses{
	\addr{label1} Факультет математичної статистики, Університет Лімерика, Лімерик V94 T9PX, Ірландія
	\addr{label2} Центр дослідження рідинних і складних систем, Університет Ковентрі, Ковентрі, CV1 5FB, Англія
}

\makeukrtitle

\begin{abstract}
	\tolerance=3000%
Складні мережі застосовувалися для кількісного порівняння міфологічних текстів вже понад десять років. Це дозволило нам виявити спільні риси між текстами в різних культурах, а також кількісно визначити важливість деяких героїчних персонажів. Аналіз повної міфології культури вимагає збору даних з багатьох окремих міфів, що займає багато часу та часто є непрактичним. У цій роботі ми намагаємося це обійти, аналізуючи мережу персонажів у словнику міфологічних персонажів. Ми показуємо, що головні персонажі, визначені різними показниками центральності, узгоджуються з основними героями ірландських саг. Хоча значна частина ірландської міфології була втрачена, ми демонструємо, що ці найбільш центральні персонажі дуже стійкі навіть до великого випадкового видалення країв вибірки.
	\keywords складні мережі, соціальні мережі, міфологія
	
\end{abstract}

\lastpage

\end{document}